\documentclass[a4paper,11pt]{article}

\usepackage{jheppub} 

\usepackage[T1]{fontenc} 
\usepackage{mathtools,slashed}

\usepackage{comment}
\usepackage{subcaption}
\usepackage{graphicx}

\title{\boldmath The Generalized Uncertainty Principle. Massless QED Renormalization}

\author[a,b]{Ezequiel Valero,}
\author[a]{Hector Gisbert}
\author[a]{and Victor Ilisie}
\affiliation[a]{Escuela de Ciencias, Ingeniería y Diseño, Universidad Europea de Valencia,\\ 
Paseo de la Alameda 7, 46010, Valencia, Spain}
\affiliation[b]{Facultat de Física, Universitat de València,\\ Carrer del Dr. Moliner, 50, 46100 Burjassot, Valencia, Spain}

\emailAdd{ezequiel.valero@universidadeuropea.es}
\emailAdd{hector.gisbert@universidadeuropea.es}
\emailAdd{victor.ilisie@universidadeuropea.es}

\abstract{The effective-field-theory interpretation of generalized uncertainty principle (GUP) deformations does not by itself establish whether their tree-level operator content is closed under renormalization. We investigate this question in massless GUP-deformed quantum electrodynamics at one loop and to first order in the deformation parameter. We compute the ultraviolet divergences of the relevant Green functions and determine the counterterm structure. Our central result is that, after eliminating redundant operators, renormalization requires only one independent physical counterterm beyond those associated with the tree-level theory: an axial–axial four-fermion interaction. The original deformation therefore admits a minimal radiative completion at this order. We derive the leading logarithmic running of the induced Wilson coefficient and discuss its operator-level correspondence with the contact interaction arising in Einstein–Cartan gravity.}

\begin{document} 
\maketitle
\flushbottom

\section{Introduction}

The Heisenberg uncertainty principle constitutes one of the fundamental ingredients of quantum mechanics and quantum field theory. In recent decades, several quantum-gravity-inspired approaches,
including string theory \cite{KONISHI1990276,Chang_2011},
black-hole physics \cite{Scardigli_1999}, and non-commutative
geometries \cite{Maggiore:1993kv,Pramanik:2013zy},
have suggested modifications of the standard uncertainty relations
associated with the emergence of a minimal observable length
\cite{Hossenfelder:2012jw} . Such modifications are commonly encoded through generalized
uncertainty principles (GUPs), formulated by deforming the canonical
commutation relations with momentum-dependent terms
\cite{Bang:2006va,Bosso:2023aht,Pedram:2011aa,Galan:2007ns,Das:2012rv,Pedram:2011gw,Mignemi:2011wh,Nozari:2012gd,Tedesco:2011iv,Balasubramanian:2014pba,Tkachuk:2013lta,Bosso:2022vlz,Pramanik:2013zy,Bosso:2018uus,Chung:2018btu,Bosso:2020fos,Bosso:2020jay,Scardigli_1999,Hossenfelder_2006,Das_2016,Maccone_2014,Tawfik_2014,Hossenfelder_2013,Nenmeli_2021,Chen_2013,Basilakos_2010,Gecim_2017,Gecim_2018,KONISHI1990276,Chang_2011,Gomes_2022,Das_2008,ali,Marin}
\begin{align}
[\hat{x}^\mu, \hat{x}^\nu]= -i\eta^{\mu\nu}(1+\beta \hat{p}^2)
+\cdots , ,
\end{align}
where $\beta=\beta_0/\Lambda^2$ introduces a new energy scale $\Lambda$ associated with the onset of minimal-length effects. While these deformations originally emerged in the context of quantum gravity, they can also be interpreted more generally as an effective-field-theory (EFT) parametrization of possible violations of the standard Heisenberg algebra at high energies.

In a previous work \cite{Valero:2025GUP}, we analyzed the generalized uncertainty principle from this EFT perspective and discussed several conceptual subtleties that frequently appear in the literature. In particular, we emphasized the distinction between the auxiliary momentum variable generating translations and the physical momentum entering observable scattering processes, clarifying how momentum conservation can consistently be implemented while avoiding issues such as the soccer-ball problem. Within this framework, we also derived new phenomenological bounds on the GUP scale from high-energy Compton scattering data, showing that present collider and precision experiments may already be sensitive to these effects.

However, once the generalized uncertainty principle is embedded into
quantum field theory, the question of quantum consistency becomes
unavoidable. The deformation introduces higher-derivative corrections
into the effective Lagrangian, which modify propagators, interaction
vertices and, consequently, the ultraviolet structure of loop
amplitudes. Although a considerable part of the literature on
GUP-inspired quantum field theories has focused on tree-level
phenomenology, comparatively less attention has been paid to their
radiative structure and renormalization. In particular, it is important
to establish whether the perturbative expansion remains consistent
beyond tree level, whether the ultraviolet divergences can be absorbed
into a finite set of local counterterms at a given order in the
deformation parameter, and whether quantum corrections generate
additional interactions not present in the classical formulation.

The purpose of the present work is to address these questions in
generalized quantum electrodynamics. We work consistently to first order
in the deformation parameter $\beta$ and analyze the massless theory at
one-loop order using dimensional regularization. Starting from the
GUP-deformed QED Lagrangian, we derive the corresponding propagators and
interaction vertices and compute the ultraviolet-divergent parts of the
relevant one-particle-irreducible Green functions. The calculation
includes the photon and fermion two-point functions, the fermion--photon
vertices with increasing photon multiplicity, and the four-fermion
sector. This provides a direct test of the closure and consistency of
the theory under radiative corrections.

At one loop, the ordinary dimension-four QED sector retains its standard
renormalization structure. No new ultraviolet divergence modifies the
photon higher-derivative two-point contribution directly, while the
divergences appearing in the fermionic Green functions can be absorbed
consistently into local counterterms. After the redundant contributions
are removed, these divergences collapse onto the same physical
higher-derivative photon direction already present in the original
theory. The corresponding gauge-parameter dependence cancels in the
physical projection, providing a non-trivial check of the calculation.

The four-fermion sector reveals a qualitatively new effect. Radiative
corrections generate an additional axial--axial contact interaction,
showing that the original GUP deformation is not by itself closed under
one-loop renormalization. Remarkably, however, only one new independent
physical interaction is required at this order. The resulting
renormalized theory therefore remains highly constrained: after the
elimination of redundant structures, its dimension-six physical content
is described by the original Lagrangian with a
single radiatively generated axial--axial four-fermion interaction.

An important aspect of our approach is that the theory is treated
strictly as an effective field theory truncated at
$\mathcal{O}(\beta)$. Within this EFT interpretation, we also determine the
renormalization-group evolution of the independent couplings.

The paper is organised as follows. In sections \ref{minimal} and \ref{modified}, we introduce the minimal GUP model and construct the complete Hermitian Lagrangian for GUP-deformed QED. In section \ref{feynrules}, we derive the corresponding propagators and interaction vertices to $\mathcal{O}(\beta)$. Section \ref{oneloopcorrections} presents the one-loop corrections in the massless limit, including the new vertex structures generated at this order. In sections \ref{counter} and \ref{fourfermion}, we determine the full set of one-loop counterterms, identify the additional operators and their Wilson coefficients, and establish the complete operator basis, including four-fermion interactions. Finally, in section \ref{rge}, we derive the one-loop renormalisation group equations.

\section{Minimal GUP model}
\label{minimal}
The generalized uncertainty principle (GUP) provides a phenomenological
framework to parametrise possible deviations from the standard
Heisenberg uncertainty relations at short distances or high energies
\cite{Kempf_1995,Bang:2006va,Pedram:2011aa,Nozari:2012gd,
Hossenfelder:2012jw,Tawfik_2014,Bosso:2023aht}.
Such modifications are motivated by arguments from string theory
\cite{KONISHI1990276,Chang_2011}, black-hole physics
\cite{Scardigli_1999,Hossenfelder:2012jw},
and non-commutative geometries
\cite{Maggiore:1993kv,Quesne:2006is,Mignemi:2011wh,Pramanik:2013zy}.
Their implications have also been explored in black-hole
thermodynamics and Hawking radiation
\cite{Basilakos_2010,Chen_2013,Gecim_2017,Gecim_2018}, which suggest the emergence
of a minimal observable length, often associated with the Planck scale.

At the level of quantum mechanics, these effects can be incorporated through deformations of the canonical commutation relations. In the relativistic covariant formulation, the standard commutation relations are given by
\begin{align}
[\hat{x}^\mu,\hat{p}^\nu]
=
-i\eta^{\mu\nu} \, ,
\qquad
[\hat{x}^\mu,\hat{x}^\nu]
=
[\hat{p}^\mu,\hat{p}^\nu]
=
0 \, ,
\end{align}
where $\eta^{\mu\nu}=\mathrm{diag}(1,-1,-1,-1)$ is the Minkowski metric. These relations lead to the usual Heisenberg uncertainty principle and constitute the starting point of ordinary quantum field theory. In this work we consider the minimal isotropic deformation of the covariant Heisenberg algebra \cite{Quesne:2006is,Todorinov:2018arx,Valero:2025GUP}, given by
\begin{align}
[\hat{x}^\mu,\hat{p}^\nu]
=
-i\eta^{\mu\nu}(1+\beta \hat{p}^2) - 2 i \beta \hat{p}^\mu \hat{p}^\nu \, ,
\label{gup_intro}
\end{align}
where $\beta=\beta_0/\Lambda^2$ introduces a new scale $\Lambda$ associated with the onset of GUP effects, while $\beta_0$ is dimensionless. The deformation parameter $\beta$ has dimensions of inverse squared energy and controls the strength of the higher-momentum corrections.

The modified commutation relation \eqref{gup_intro} preserves the commutativity of both coordinates and momenta, avoiding the introduction of non-commutative spacetime structures and the additional complications they entail. The deformation can be interpreted as the leading-order contribution of an effective expansion in powers of momentum over the new-physics scale $\Lambda$.

In order to consistently quantize the theory, it is convenient to introduce an auxiliary momentum operator $\hat{k}^\mu$ satisfying the standard canonical algebra
\begin{align}
[\hat{x}^\mu,\hat{k}^\nu]
=
-i\eta^{\mu\nu} \, .
\end{align}
The physical momentum operator $\hat{p}^\mu$ can then be expressed perturbatively in terms of $\hat{k}^\mu$ as
\begin{align}
\hat{p}^\mu
=
\hat{k}^\mu(1+\beta \hat{k}^2)
+
\mathcal{O}(\beta^2) \, .
\end{align}
Within this framework, the auxiliary quantity $k^\mu$ generates translations in the usual way, while the physical momentum $p^\mu$ corresponds to the measurable momentum entering scattering processes and satisfying the standard momentum-conservation laws.

Expressing the momentum operators in terms of derivatives, the previous relation induces the replacement
\begin{align}
i\partial^\mu
\rightarrow
i\partial^\mu(1-\beta \square)
+
\mathcal{O}(\beta^2) \, ,
\label{subst_intro}
\end{align}
where $\square=\partial^\mu\partial_\mu$ is the d'Alembertian operator. Consequently, the generalized uncertainty principle generates higher-derivative operators in the effective Lagrangian, leading to modified propagators and interaction vertices once gauge interactions are introduced.

Throughout this work the theory will be treated strictly as an effective field theory truncated at first order in $\beta$. Accordingly, all propagators, interaction terms and loop corrections will consistently be expanded up to $\mathcal{O}(\beta)$, with the higher-derivative contributions interpreted as EFT corrections suppressed by the scale $\Lambda$.

\section{Modified QED Lagrangian}
\label{modified}

In order to obtain the modified QED interaction terms and propagators, let us start by considering
the free hermitic Dirac and photon field Lagrangian i.e., 
\begin{align}
    \mathcal{L} &= \frac{i}{2} \bar{\psi} \gamma^\mu \overset{\leftrightarrow}{\partial}_\mu \psi - m \bar{\psi} \psi - \frac{1}{4} F^{\mu\nu}F_{\mu\nu} \notag \\ &= \frac{i}{2} \left( \bar{\psi} \gamma^\mu \partial_\mu \psi - (\partial_\mu \bar{\psi}) \gamma^\mu \psi \right) - m \bar{\psi} \psi - \frac{1}{4} F^{\mu\nu}F_{\mu\nu} \, .
\end{align}
After performing the $\partial_\mu \to \partial_\mu (1-\beta \partial^\rho \partial_\rho)$ shift, and switching the ordinary partial derivative with the covariant derivative, one obtains the expression of the GUP modified QED Lagrangian that includes the extra interaction terms and modified propagators
\begin{align}
\label{QEDG}
\mathcal{L}^{\text{GUP}}_{\text{QED}} &= \bar{\psi} \Big(i\slashed{\partial}(1-\beta \square) - m \Big)\psi  - \frac{1}{4} F^{\mu\nu}F_{\mu\nu} + \frac{\beta}{2} F^{\mu\nu} \Big(\square F_{\mu\nu} \Big) \notag \\
 & \qquad \qquad \qquad -eQA_\mu \bar\psi \gamma^\mu \psi  + {\mathcal{L}}_{I}^\beta + \mathcal{L}_{\xi} + \mathcal{O}(\beta^2)  \, .
\end{align}
where we have introduced the usual short hand notation:
\begin{align}
\slashed{\partial} \equiv \gamma^\mu \partial_\mu \, , \qquad \qquad \square \equiv \partial^\alpha \partial_\alpha \, ,
\end{align}
and where ${\mathcal{L}}_{I}^\beta$ is the $\mathcal{O}(\beta)$ complete interaction Lagrangian\footnote{Maintaing all terms, up tp a total derivative, including terms that have been neglected in \cite[Valero:2025GUP] that vanish on-shell i.e, without applying the equations of motion (EOM).} containing the newly generated interaction terms. It is given by
\begin{align}
{\mathcal{L}}_{I}^\beta &= \beta\, e Q\, \bar{\psi} \gamma^\mu \Big[
\frac{1}{2}A_{\mu}\Box
+\frac{1}{2} \overleftarrow{\Box}A_{\mu}
+\left(\partial_{\mu}A^{\rho}\right)\partial_{\rho}
+\overleftarrow{\partial}_{\rho}
 \left(\partial_{\mu}A^{\rho}\right)
+A^{\rho}\partial_{\rho}\partial_{\mu}
+\overleftarrow{\partial}_{\mu}
 \overleftarrow{\partial}_{\rho}A^{\rho} \notag \\ & \qquad\qquad\qquad
+(\partial_{\mu}\partial_{\rho}A^{\rho})
+\frac{1}{2}\left(\partial_{\rho}A^{\rho}\right)\partial_{\mu}
+\frac{1}{2}\overleftarrow{\partial}_{\mu}
 \left(\partial_{\rho}A^{\rho}\right)
 \Big]\psi \notag \\
&\quad \quad 
+ \frac{i}{2} \beta\, (eQ)^2\, \bar{\psi} \gamma^\mu \Big[
  A^\rho A_\rho\, \partial_\mu 
  - \overset{\leftarrow}{\partial}_\mu A^\rho A_\rho 
  + 2 A_\mu A^\rho\, \partial_\rho 
  - 2 \overset{\leftarrow}{\partial}_\rho A_\mu A^\rho
\Big] \psi \notag \\
&\quad \quad \quad \quad 
- \beta\, (eQ)^3\, \bar{\psi} \gamma^\mu A_\mu A^\rho A_\rho\, \psi\, .
\label{LIBB}  
\end{align}
which is the interaction Lagrangian that we will be using for our calculations. Next, we shall derive the Feynman rules and check the asociated Ward indenties related to the full Lagrangian.

\section{Feynman Rules}
\label{feynrules}

\begin{figure}[t!]
\centering
\includegraphics[scale=0.52]{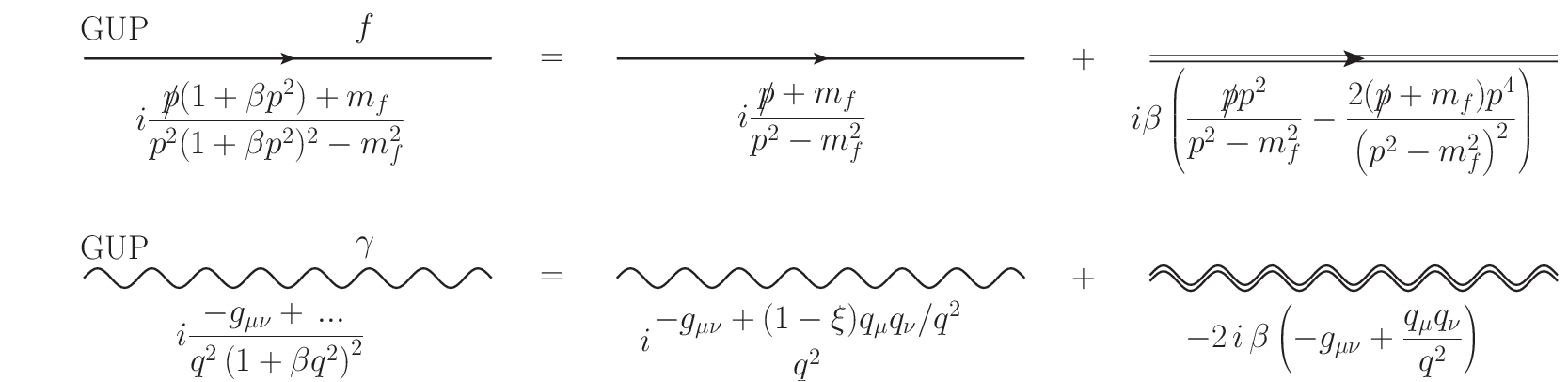}
\caption{Modified Feynman rules for the propagators expressed as a sum of the SM contributions and the $\mathcal{O}(\beta)$ corrections.} 
\label{prop}
\end{figure}

\begin{figure}[t!]
\centering
\includegraphics[scale=0.4]{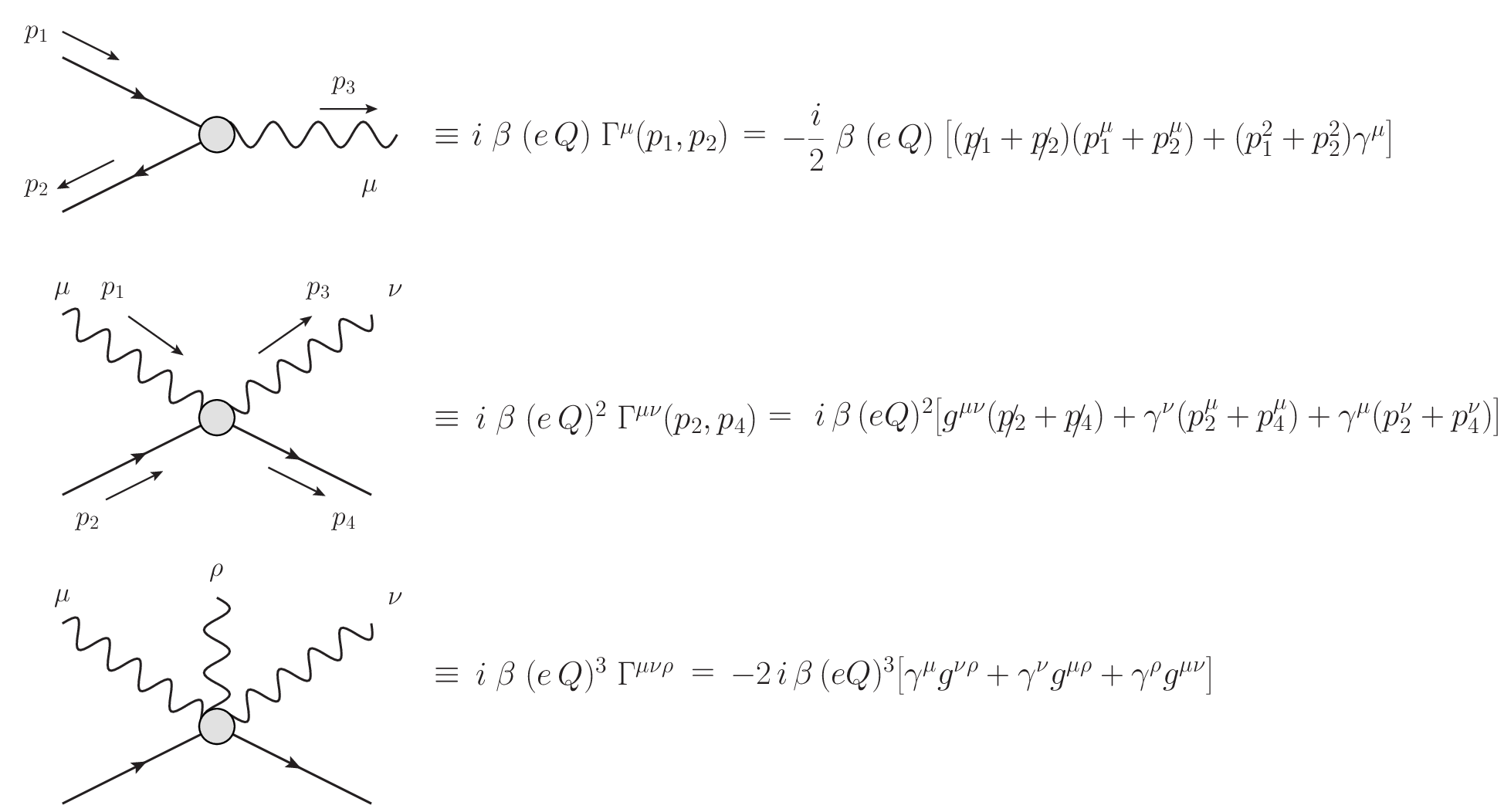}
\caption{Feynman rules of the new QED vertices at $\mathcal{O}(\beta)$ where the momentum flow is given by the corresponding arrows. Momentum conservation reads $p_1=p_2+p_3$ for the 3-point vertex and, $p_1+p_2=p_3+p_4$ for the 4-point vertex.} 
\label{vert}
\end{figure}

In order to find the Feynman propagator of the photon field, we shall employ the usual gauge-fixing term given by:
\begin{align}
\mathcal{L}_{\xi} = - \frac{1}{2\xi} \Big(\partial_\mu A^\mu \Big)^2 \, ,
\end{align}
and so, we obtain the following equation of motion for the photon field
\begin{align}
\left( g^{\mu\nu}\square(1-\beta\square)^2 + \frac{1}{\xi}\partial^\mu\partial^\nu  - \partial^\mu\partial^\nu (1-\beta\square)^2 \right) A_\nu \equiv O^{\mu\nu}_\xi A_\nu = 0 \, .
\end{align}
The Feynman propagator $iD^F_{\mu\nu}$ must then satisfy
\begin{align}
O^{\mu\nu}_\xi \, iD^F_{\nu\alpha}(x) = i \, \delta^\mu_\alpha \delta^{(4)}(x) \, . 
\end{align}
which at $\mathcal{O}(\beta)$ gives (ignoring the $i\epsilon$ term):
\begin{align}
iD^F_{\mu\nu} (x) = i \int \frac{d^4p}{(2\pi)^4} D^F_{\mu\nu}(q^2) e^{-iqx}  \, , 
\end{align}
with the expression of the propagator $D^F_{\mu\nu} (q^2)$, in momentum space, given by
\begin{align}
iD^F_{\mu\nu} (q^2) = i \frac{ -{g}_{\mu  \nu } + (1-\xi ) {q}_{\mu} {q}_{\nu} / {q}^2}{{q}^2} 
- 2i\beta \left( -{g}_{\mu \nu}  +\frac{q_\mu q_\nu}{q^2} \right)   
\, .
\label{tree_lev_ph}
\end{align}
On the other hand, the fermionic Feynman propagator $S_F(x)$ must satisfy the modified equation
\begin{align}
iS_F(x) = \Big( i \slashed{\partial} (1-\beta\square) + m  \Big) \Delta_F(x) \, ,
\end{align}
and so, we can easily find the expression for the propagator to be
\begin{align}
iS_F(p) = \frac{i\left(\slashed{p}(1 + \beta p^2) + m \right)}{p^2(1 + \beta p^2)^2 - m^2}
= \frac{i(\slashed{p} + m)}{p^2 - m^2} 
+ i\beta \left[
    \frac{\slashed{p} \, p^2}{p^2 - m^2}
    - \frac{2p^4(\slashed{p} + m)}{(p^2 - m^2)^2}
\right] 
+ \mathcal{O}(\beta^2) \,.
\end{align}
The Feynman rules for the modified propagators are shown in Figure~\ref{prop}, where we have split the full propagators
(labeled as GUP) into the {\it ordinary} Standard Model contributions and, the new $\mathcal{O}(\beta)$ contributions. Finally, the new vertices at $\mathcal{O}(\beta)$ corresponding to the interaction terms can be deduced following standard procedures and they are given in Figure~\ref{vert}.

\section{Tree-Level Ward Identities}
\label{ward}

The tree-level Ward indentity for the three-point vertex, using the momentum distribution given in Figure~\ref{vert}, i.e., $p_1-p_2=p_3$, can be expressed as:
\begin{align}
p_{3,\mu} \, \Gamma^\mu_T(p_1,p_2) = S_F^{-1}(p_2) -  S_F^{-1}(p_1) \, , 
\end{align}
where we have considered the total vertex $\Gamma_T^\mu$ given by the SM usual expression and the $\beta$ contribution (except for the $ieQ$ factor), i.e., 
\begin{align}
 \Gamma_T^\mu(p_1,p_2) =  -\gamma^\mu + \Gamma^\mu (p_1,p_2) = - \gamma^\mu - \frac{1}{2} \beta \Big[ \gamma^\mu (p_1^2 + p_2^2) + 
(\slashed{p_1} + \slashed{p_2}) (p_1^\mu + p_2^\mu)  \Big] \, .
\end{align} 
After some algebra, one can check that the previous Ward indentity holds true. In the massless limit it can been seen in a straightforward manner. On one hand
\begin{align}
p_{3,\mu} \Gamma^\mu_T &= - \slashed{p_3} - \frac{\beta}{2}  \Big[ \slashed{p_3} (p_1^2 + p_2^2) + 
(\slashed{p_1} + \slashed{p_2}) (p_1\cdot p_3 + p_2 \cdot p_3)  \Big]  \notag \\
 & = 
\slashed{p_2} (1+\beta p_2^2) - \slashed{p_1} (1+\beta p_1^2) \, .
\label{wardGamma1}
\end{align}
where we have used momentum conservation to eliminate $p_3$. On the other hand, the complete inverse massless fermionic propagator reads
\begin{align}
S_F^{-1}(p) &= \left[ \frac{\slashed{p}}{p^2} - \beta \slashed{p} \right]^{-1} = \left[ \frac{\slashed{p} (1-\beta p^2)}{p^2} \right]^{-1}  = \left[ \frac{(1-\beta p^2)}{\slashed{p}} \right]^{-1} \notag \\ & = \frac{\slashed{p}}{(1-\beta p^2)} =  \slashed{p}(1+\beta p^2) + \mathcal{O}(\beta^2) \, ,
\end{align}
and so, $S_F^{-1}(p_2) -  S_F^{-1}(p_1)$ gives the same result as \eqref{wardGamma1}. As for the two-photon vertex, we have the following Ward indentities
\begin{align}
p_{1,\mu} \Gamma^{\mu\nu}(p_2,p_4) &= \Gamma^\nu(p_2,p_4-p_1) -  \Gamma^\nu(p_2+p_1,p_4)  \, ,
 \notag  \\ 
p_{3,\nu} \Gamma^{\mu\nu}(p_2,p_4) &= \Gamma^\mu(p_2-p_3,p_4) -  \Gamma^\mu(p_2,p_4+p_3) \, , 
\label{wardGamma2}
\end{align}
which is straightfowrward to check using momentum conservation $p_1+p_2=p_3+p_4$.

Finally, for the three-pont vertex, considering $k_1^\mu, \, k_2^\nu$ and $k_3^\rho$ the four-momenta of the (incoming) photons and $p$ and $p'$ the incoming momentum of the fermion and the outgoing momentum of the anti-fermion, we have the following
\begin{align}
k_{1,\mu} \Gamma^{\mu\nu\rho}  &= \Gamma^{\nu\rho}(p,p'-k_1) - \Gamma^{\nu\rho}(p+k1,p')  \, ,
 \notag  \\ 
k_{2,\nu} \Gamma^{\mu\nu\rho}  &= \Gamma^{\mu\rho}(p,p'-k_2) - \Gamma^{\mu\rho}(p+k2,p') \, , 
 \notag  \\ 
k_{3,\rho} \Gamma^{\mu\nu\rho} &= \Gamma^{\mu\nu}(p,p'-k_3) - \Gamma^{\mu\nu}(p+k3,p') \, . 
\label{wardGamma3}
\end{align}

\section{Massless QED one-loop corrections}
\label{oneloopcorrections}

In the following we will use dimensional regularization with $D=4+2\epsilon$ the number of space-time dimensions (with $\epsilon<0$ for UV divergences) and where, we introduce the short-hand notation
\begin{align}
    \frac{1}{\hat{\epsilon}} = \frac{1}{\epsilon} + \gamma_E - \ln(4\pi) \, .
\end{align}
Also, as we are interested in the renormalizability of the theory and its behaviour in the UV limit and its renomalizability, we shall only keep the UV-divergent poles.

\subsection{Photon self-energy and Dyson summation}

\begin{figure}[t!] 
\centering
\includegraphics[scale=0.3]{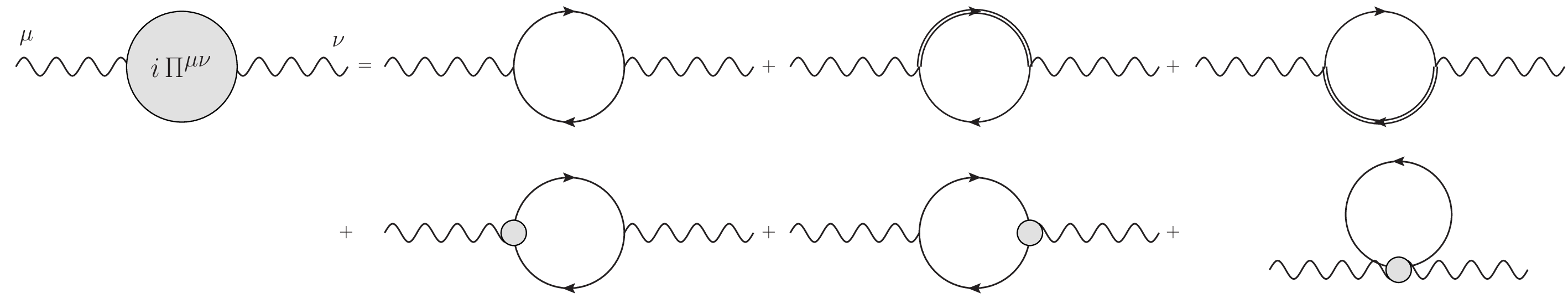}
\caption{SM and GUP contributions at $\mathcal{O}(\beta)$ to the photon self-energy.} 
\label{ph_self}
\end{figure}

The one-loop photon self-energy diagrams are shown in Figure~\ref{ph_self}, where the first diagram stands for the SM contribution. We can write this self-energy contribution as a sum of the finite (leading log) part (labeled as $f$) and a UV-divergent part (labeled as $\epsilon$)
\begin{align}
i \Pi^{\mu\nu} = i\Pi^{\mu\nu}_f + i\Pi^{\mu\nu}_\epsilon \, .
\end{align}
The finite parts can be separated in terms of the SM contribution (labeled as SM), and the new contributions (labeled as $\beta$). In the massless fermion limit the UV divergent term $\Pi^{\beta}_\epsilon$ is zero, as expected, therefore we obtain
\begin{align}
i \Pi^{\mu\nu} = i(-g^{\mu\nu} q^2 + q^\mu q^\nu ) \Big( \Pi^{\text{SM}}_f + \Pi^{\text{SM}}_\epsilon \Big) \, ,
\end{align}
with the following explicit expression in the $\overline{\text{MS}}$ scheme
\begin{align}
\Pi^{\text{SM}}_\epsilon &= - Q^2 \left(\frac{\alpha}{4\pi}\right) \frac{4}{3\hat{\epsilon}} \mu^{2\epsilon}  \, .
\end{align}
As it can be inferred from the previous result, it turns out that there are no UV-divergent parts proportional to $\beta$ for the self-energy in the massless limit. It also turns out that, in this limit, there are no finite corrections either, therefore $\Pi^{\beta}_f=0 =\Pi^{\beta}_\epsilon$. One might naively conclude that, since the one-loop self-energy contains no $\beta$-dependent terms, the propagator will not receive any GUP corrections at one loop. However, this conclusion is incorrect, as becomes clear once we perform the Dyson resummation. The complete, one-loop corrected and renormalized $D_{\mu\nu}^{F(1)}$ propagator will be given by
\begin{align}
i D_{\mu\nu}^{F(1)} = i D_{\mu\nu}^{F} + i D_{\mu\alpha}^{F} \left(i\Pi^{\alpha\beta}_f \right) i D_{\beta\nu}^{F} + ...  \, , 
\end{align}
where $i D_{\mu\nu}^{F}$ is the tree-level propagator \eqref{tree_lev_ph} and with $i\Pi^{\mu\nu}_f$, the finite, renormalized self energy function. Explicitly
\begin{align}
i D_{\mu\nu}^{F(1)} &= i D_{\mu\nu}^{F} + i\, D_{\mu\alpha}^{F} (-g^{\alpha\beta}q^2 + q^\alpha q^\beta) \Big( i \Pi^{\text{SM}}_f \Big) i D_{\beta\nu}^{F} + ...  
\notag \\
&= \frac{i}{q^2} \Big( -{g}_{\mu  \nu } + \frac{q_\mu q_\nu}{q^2} \Big) \left(1-\Pi_f^{\text{SM}} \right)- i \xi \frac{q_\mu q_\nu}{q^4} 
\notag \\ 
& \qquad \qquad - 2i\beta \Big( -{g}_{\mu \nu}  +\frac{q_\mu q_\nu}{q^2} \Big) \Big(1-2\Pi^{\text{SM}}_f \Big)  
+ ...  \, .
\end{align}
After adequately grouping terms, we obtain
\begin{align}
i D_{\mu\nu}^{F(1)} &= i\Big( -{g}_{\mu  \nu } + \frac{q_\mu q_\nu}{q^2} \Big) \left( \frac{1}{q^2(1+\Pi_f^{\text{SM}})} -\frac{2\beta}{1+2\Pi^{\text{SM}}_f } \right)- i \xi \frac{q_\mu q_\nu}{q^4}  \, . 
\end{align}
Indeed, one can observe that the SM self-energy directly contributes to the one-loop $\beta$ correction of the photon propagator.

\subsection{Fermion self-energy and Dyson summation}

\begin{figure}[t!]
\centering
\includegraphics[scale=0.33]{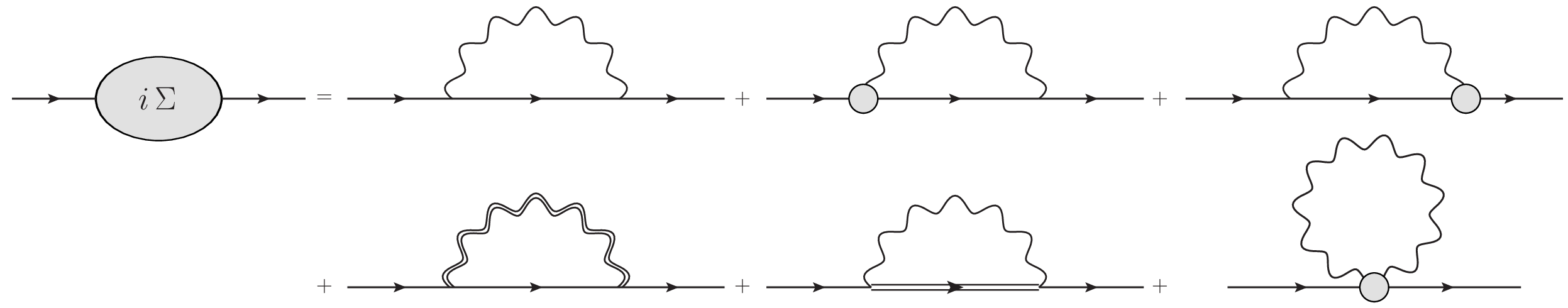}
\caption{SM and GUP contributions at $\mathcal{O}(\beta)$ to the fermion self-energy.} 
\label{fer_self}
\end{figure}

The one-loop fermion self-energy is given in Figure~\ref{fer_self}, where, the first diagram stands for the SM contribution. We shall separate the self-energy as
\begin{align}
i \Sigma = i \Sigma_{\text{SM}}(p) + i \Sigma_{\beta}(p)  = i \slashed{p} \, \tilde\Sigma_{\text{SM}}  + i \beta p^2  \slashed{p} \, \tilde\Sigma_{\beta} \, ,    
\end{align}
The explicit SM expressions of the UV poles are given by
\begin{align}
\Sigma_{\text{SM}}^\epsilon &= - Q^2 \left(\frac{ \alpha }{4 \pi}\right) \frac{\xi}{\hat{\epsilon}} \mu^{2 \epsilon} \, , \quad \Sigma_\beta^\epsilon =Q^2  \, \left(\frac{\alpha}{4 \pi }\right) \frac{ (5 - 4 \xi) } {2 \,  \hat{\epsilon} } \,  \mu ^{2 {\epsilon} } \, .
\end{align}
The complete propagator, including one-loop contributions $i S_F^{(1)}$, will be given by the Dyson summation. Considering the renormalized self-energy $i\Sigma_f$, obtained by summing the corresponding counter-terms we have
\begin{align}
i S_F^{(1)} = i S_{F}  + i S_{F} \big(i \Sigma_f \big) i S_F + ...  \, , 
\end{align}
where $i S_{F}$ is the tree-level massless fermion propagator. Explicitly the previous expression reads
\begin{align}
i S_F^{(1)} = i\left(\frac{\slashed{p}}{p^2} - \beta \slashed{p}\right) + i\left(\frac{\slashed{p}}{p^2} - \beta \slashed{p} \right) \big(i \Sigma \big) \, i\left(\frac{\slashed{p}}{p^2} - \beta \slashed{p} \right) + ... \, .  
\end{align}
After adequately grouping terms, we obtain
\begin{align}
i S_F^{(1)} = i\frac{\slashed{p}}{p^2 (1+\Sigma_{\text{SM}}^f)} - i \beta \frac{\slashed{p}}{1+2\Sigma_{\text{SM}}^f - \Sigma_\beta^f} \, .
\end{align}

\subsection{Three-point vertex correction}

\begin{figure}[t!]
\centering
\includegraphics[scale=0.33]{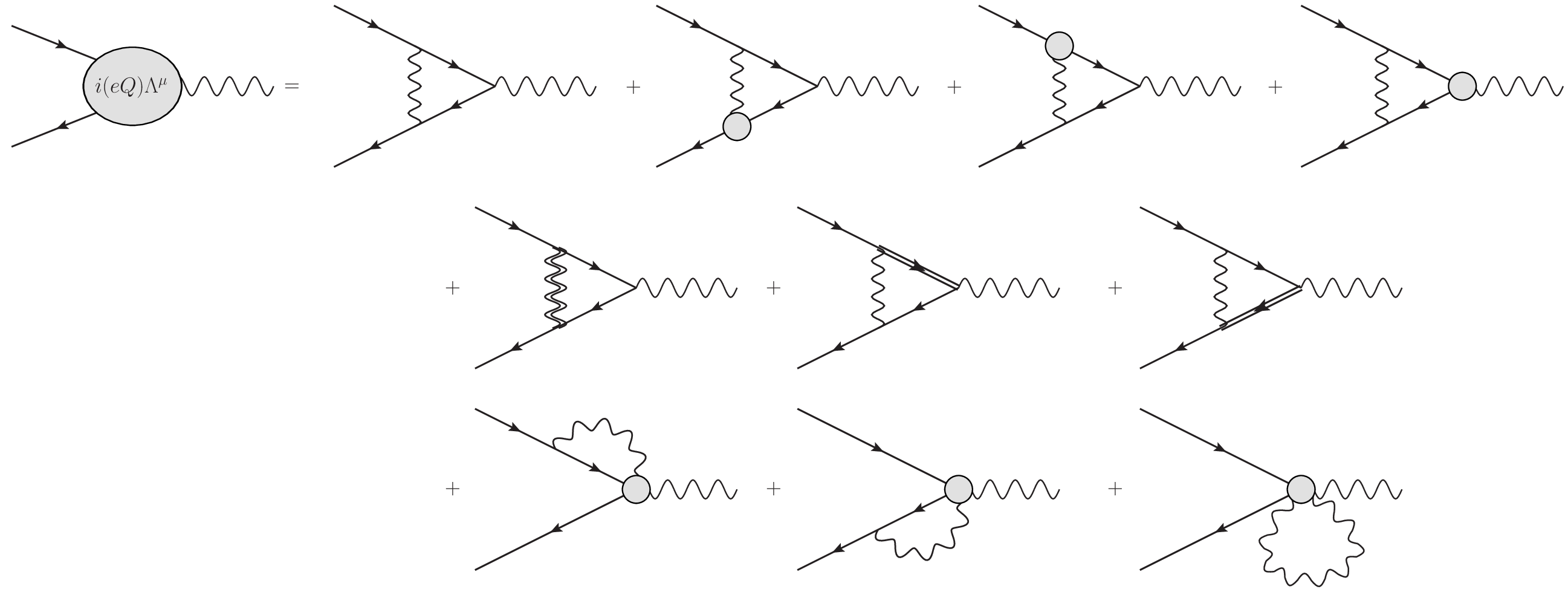}
\caption{SM and GUP one-loop contributions at $\mathcal{O}(\beta)$ to the three-point vertex.} 
\label{ver_cor}
\end{figure}

The one-loop contributions to the three-point vertex correction are given in Figure~\ref{ver_cor}, where, the first diagram stands for the SM contribution. The complete one-loop corrected vertex will be given by
\begin{align}
i (eQ)\Gamma^\mu_{(1)} = -i (eQ) \gamma^\mu \Big(1+\Lambda_{\text{SM}} \Big) + i \beta (eQ) \Gamma^\mu \Big(1+\Delta_{\Gamma}\Big) + i \beta (eQ) \tilde{\Lambda}^\mu \, ,    
\end{align}
where $\tilde{\Lambda}^\mu$ are the corrections that do not factorize with the tree-level $\mathcal{O}(\beta)$ contribution to the three-point function given by $\Gamma^\mu=\Gamma^\mu(p_1,p_2)$. Thus, we can write the one-loop correction as:
\begin{align}
i (e Q) \Lambda^\mu &= -i (e Q) \Lambda^\mu_{\text{SM}} + i \beta (e Q) \Lambda^\mu_{\beta}
\notag \\ &= -i (eQ)\gamma^\mu\Lambda_{\text{SM}} + i \beta (eQ) \Big(\Gamma^\mu  \Delta_{\Gamma} + \tilde{\Lambda}^\mu \Big).
\end{align}
The SM UV-divergent correction is given by 
\begin{align}
 \Lambda_{\text{SM}}^\epsilon = -Q^2  \left( \frac{\alpha}{4\pi} \right) 
\frac{\xi}{\hat{\epsilon}} \mu^{2\epsilon} \, .
\end{align}
The one-loop $\beta$ contributions are given by $\Delta_\Gamma$ and $\tilde{\Lambda}^\mu$. They  are conveniently decomposed such that the coefficient $c_\Gamma$ is fixed by the longitudinal Ward--Takahashi identity, whereas $\widetilde{\Lambda}^\mu$ contains the genuinely transverse UV structures, i.e., it satisfies $p_{3,\mu} \, \tilde{\Lambda}^\mu=0$, and therefore parametrizes the possible mixing with additional effective operators. The UV divergent parts are
\begin{align}
\Delta_\Gamma^\epsilon &= Q^2  \left( \frac{\alpha}{4\pi} \right) \frac{(5-4\xi)}{2\hat{\epsilon}} \mu^{2\epsilon} \, , \notag \\[1.5ex]
\tilde{\Lambda}^\mu_\epsilon &= Q^2  \left( \frac{\alpha}{4\pi} \right) \frac{1}{\hat{\epsilon}} \mu^{2\epsilon} \, \left[ \frac{13+6\xi}{12} T_1^\mu 
+ \left( \xi - \frac{1}{2} \right) T^\mu_2  \right] \, ,
\label{3pointCorr}
\end{align}
where the auxiliary functions $T^\mu_1=T^\mu_1(p_1,p_2)$ and $T^\mu_2=T^\mu_2(p_1,p_2)$ are independently transverse, i.e., they satisfy $p_{3,\mu} \, T^\mu_{1,2}=0$, and they are given by
\begin{align}
 T_1^\mu(p_1,p_2) &= {\gamma }^{\mu} (p_2-p_1)^2 - (p_2^\mu-p_1^\mu) (\slashed{p_2} - \slashed{p_1})
 \, , \notag \\  
 T_2^\mu(p_1,p_2) &= - {\gamma }^{\mu } \left(  p_1 \cdot p_2\right) +  {\gamma }^{\mu }  \slashed{p_1}  \slashed{p_2} +  p_2^\mu \slashed{p_1} - p_1^\mu\slashed{p_2}
\, .
\end{align}
Again, one can check that the results obey the one-loop Ward indenty, i.e., 
\begin{align}
-p_{3,\mu} \, \Lambda_{\text{SM}}^\mu = \Sigma_{\text{SM}}(p_2) - \Sigma_{\text{SM}}(p_1) \, ,
\end{align}
for the SM, as usual, and
\begin{align}
p_{3,\mu} \Lambda^\mu_\beta = p_{3,\mu} \Big( \Gamma^\mu \Delta_{\Gamma} +  \tilde{\Lambda}^\mu \Big) = p_{3,\mu} \, \Gamma^\mu \Delta_{\Gamma} = \Sigma_\beta(p_2) - \Sigma_\beta(p_1) \, , 
\end{align}
for the one-loop beta correction.

\subsection{Four-point vertex correction}

\begin{figure}[t!]
\centering
\includegraphics[scale=0.42]{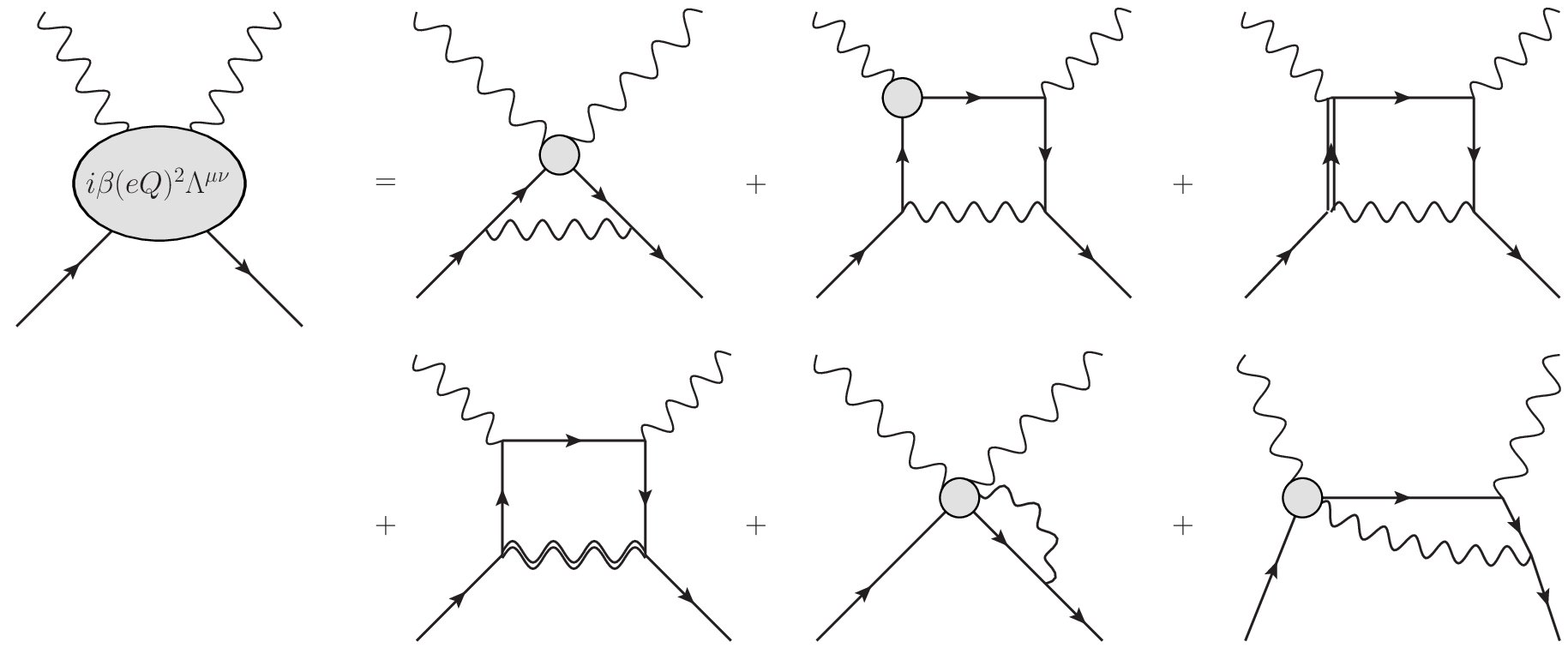}
\caption{Typical one-loop GUP contributions at $\mathcal{O}(\beta)$ to the four-point vertex.} 
\label{ver4_cor}
\end{figure}

The topological classes for the one-loop contributions to the four-point vertex correction are given in Figure~\ref{ver4_cor}. Here we haven't included the SM contribution, as it is finite and we are only interested in the UV divergent behaviour. The complete one-loop corrected vertex will be given by
\begin{align}
i \beta (eQ)^2 \Gamma^{\mu\nu}_{(1)} = i \beta (eQ)^2 \Gamma^{\mu\nu} \Big(1+\delta_{\Gamma}\Big) + i \beta (eQ)^2 \tilde{\Lambda}^{\mu\nu} \, ,    
\end{align}
where $\tilde{\Lambda}^{\mu\nu}$ are the corrections that do not factorize with the tree-level $\mathcal{O}(\beta)$ contribution to the four-point function given by $\Gamma^{\mu\nu}=\Gamma^{\mu\nu}(p_2,p_4)$. Thus, we can write the one-loop correction as:
\begin{align}
i \beta (eQ)^2 \Lambda^{\mu\nu} &= i \beta (eQ)^2 \Big(\Gamma^{\mu\nu}  \delta_{\Gamma} + \tilde{\Lambda}^{\mu\nu} \Big).
\end{align}
The one-loop UV-divergent $\beta$ contributions given by $\delta_\Gamma^\epsilon$ and $\tilde{\Lambda}^{\mu\nu}_\epsilon$ read 
\begin{align}
\delta_\Gamma^\epsilon &= Q^2  \left( \frac{\alpha}{4\pi} \right) \frac{(5-4\xi)}{2\hat{\epsilon}} \mu^{2\epsilon} \, , \notag \\[1.5ex]
\tilde{\Lambda}^{\mu\nu}_\epsilon &= Q^2  \left( \frac{\alpha}{4\pi} \right) \frac{1}{\hat{\epsilon}} \mu^{2\epsilon} \, \left( \xi - \frac{1}{2} \right) \, G^{\mu\nu}   \, ,
\label{4pointCorr}
\end{align}
where the auxiliary function $G^{\mu\nu}=G^{\mu\nu}(p_2,p_4)$ satisfies $(p_2+p_4)_\mu G^{\mu\nu} = 0 = (p_2+p_4)_\nu G^{\mu\nu} $ and it is given by
\begin{align}
G^{\mu\nu}(p_2,p_4) = g^{\mu\nu} (\slashed{p_2}+\slashed{p_4}) - \gamma^\mu (p_2^\nu + p_4^\nu) - \gamma^\nu (p_2^\mu + p_4^\mu) +  \gamma^\mu(\slashed{p_2}+\slashed{p_4})\gamma^\nu \, . 
\end{align}
Note that $\delta_\Gamma^\epsilon = \Delta_\Gamma^\epsilon$, i.e. the three- and four-point vertices receive the same factorising correction at one loop. Moreover, the one-loop coefficient of $G^{\mu\nu}$ coincides with that multiplying $T^\mu_2$. As we shall see, these relations provide a first indication of the operator structure underlying the complete one-loop counterterm basis of GUP-deformed QED.

\subsection{Five-point vertex correction}

The topological classes for the one-loop contributions to the five-point vertex correction are given in Figure~\ref{ver4_cor}. Here we haven't included the SM contribution either, as it is also finite. The complete one-loop corrected vertex will be given by
\begin{align}
i \beta (eQ)^3 \Gamma^{\mu\nu\rho}_{(1)} = i \beta (eQ)^3 \Gamma^{\mu\nu\rho} \Big(1+C_{\Gamma}\Big) \,.    
\end{align}
\begin{figure}[t!]
\centering
\includegraphics[scale=0.42]{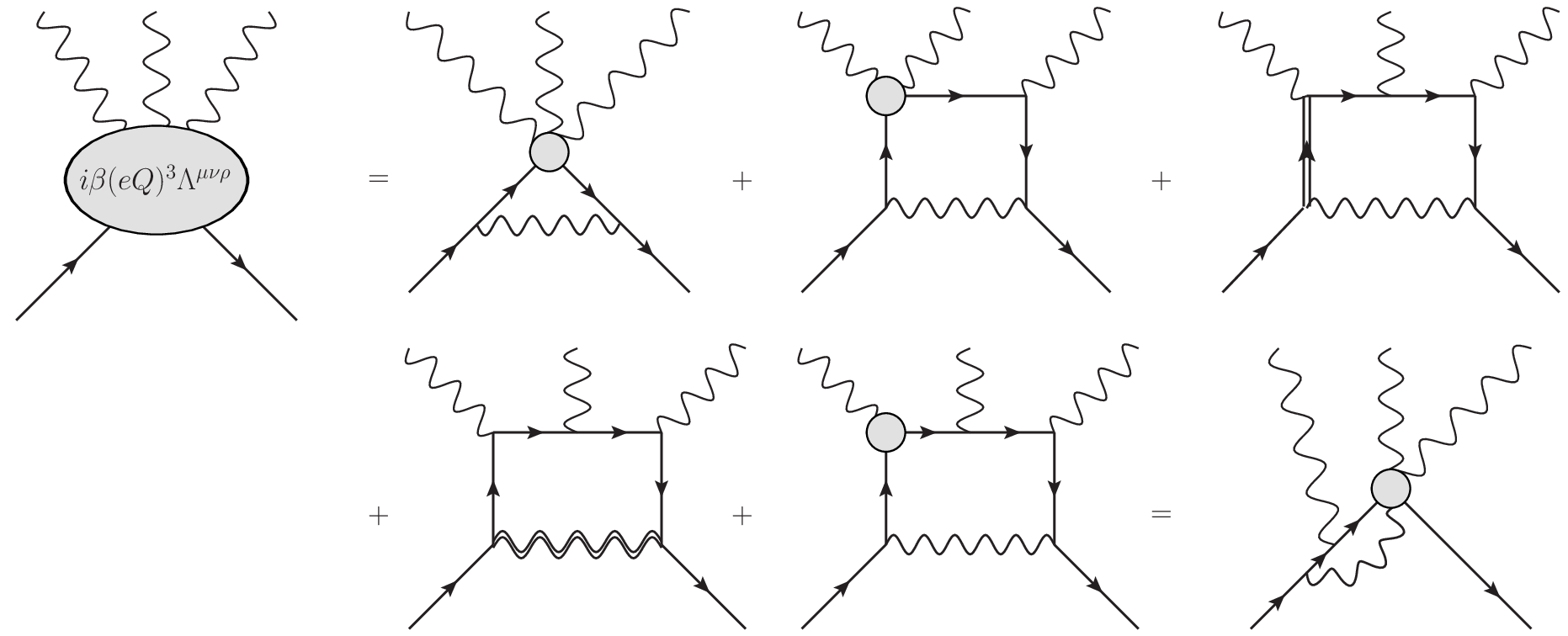}
\caption{Typical one-loop GUP contributions at $\mathcal{O}(\beta)$ to the five-point vertex.} 
\label{ver5_cor}
\end{figure} 
The one-loop UV-divergent $\beta$ contributions given by $C_\Gamma^\epsilon$ read 
\begin{align}
C_\Gamma^\epsilon &= Q^2  \left( \frac{\alpha}{4\pi} \right) \frac{(5-4\xi)}{2\hat{\epsilon}} \mu^{2\epsilon} \, , 
\label{5pointCorr}
\end{align}
Note, again, that $C_\Gamma^\epsilon = \delta_\Gamma^\epsilon = \Delta_\Gamma^\epsilon$ (the three-point, the four-point and the five-point verteces receive the same factorizing correction at one loop).

\section{Massless QED one-loop counterterm structure}
\label{counter}

The explicit one-loop calculation shows that the ultraviolet
divergences generated by a single insertion of $\mathcal{O}({\beta})$
cannot be absorbed into $\mathcal{O}_{\beta\psi}$ alone. Two additional
local, gauge-invariant structures are required. A convenient choice is
\begin{align}
\mathcal{L}_{DF}  + \mathcal{L}_{3\gamma} \equiv \beta \left(\mathcal{O}_{DF} +  \mathcal{O}_{3\gamma} \right)  \, .
\end{align}
with the additional operators given by
\begin{equation}
\mathcal{O}_{DF}
=
-(eQ)
\left(\bar{\psi}\gamma_{\mu}\psi\right)
\partial_{\nu}F^{\nu\mu}
\label{eq:ODF}
\end{equation}
and
\begin{equation}
\mathcal{O}_{3\gamma}
=
-\frac{(eQ)}{4}\,
\bar{\psi}\gamma^{\mu\nu\rho}
F_{\mu\nu}
\overleftrightarrow{D}_{\rho}\psi
\label{eq:O3gamma}
\end{equation}
where we have introduced the following notation
\begin{equation}
\gamma^{\mu\nu\rho}
=
\gamma^{[\mu}\gamma^{\nu}\gamma^{\rho]},
\qquad
\bar{\psi}\overleftrightarrow{D}_{\rho}\psi
=
\bar{\psi}D_{\rho}\psi
-
\left(\bar{\psi}\overleftarrow{D}_{\rho}\right)\psi .
\end{equation}
Their three-field Feynman rules can be normalized as
\begin{equation}
i V_{DF}^{\mu}
=
i (eQ)\,T_1^{\mu},
\qquad
T_1^{\mu}
=
q^2\gamma^{\mu}
-
q^{\mu}\slashed{q},
\end{equation}
and
\begin{equation}
i V_{3\gamma}^{\mu}
=
i (eQ)\,T_2^{\mu},
\end{equation}
with
\begin{equation}
T_2^{\mu}
=
\gamma^{\mu}\slashed{p}_1\slashed{p}_2
-
p_1^{\mu}\slashed{p}_2
+
p_2^{\mu}\slashed{p}_1
-
\left(p_1\!\cdot p_2\right)\gamma^{\mu}
=
\gamma^{\mu\alpha\beta}
p_{1\alpha}p_{2\beta}.
\end{equation}
Both structures are manifestly transverse (with $q=p_2-p_1 $, the four-momentum of the incoming photon),
\begin{equation}
q_{\mu}T_1^{\mu}=0,
\qquad
q_{\mu}T_2^{\mu}=0.
\end{equation}
The operator $\mathcal{O}_{3\gamma}$ also generate a four-field
$\bar{\psi}\psi A A$ contact interaction with the corresponding Feynman rule given by
\begin{align}
iV^{\mu\nu}_{3\gamma} = i (eQ)^2 \gamma^{\mu\nu\rho}(q_2-q_1)_\rho \, ,
\end{align}
with $q_1^\mu$ and $q_2^\nu$ the in-coming four-momenta of the photons. On the other hand, neither
$\mathcal{O}_{DF}$ nor $\mathcal{O}_{3\gamma}$ generates a $\bar{\psi}\psi A A A$ vertex.

Collecting the ultraviolet poles obtained from the two-, three-,
four- and five-field one-particle-irreducible Green functions, the
divergence generated by one insertion of $\mathcal{O}({\beta})$
can be cancelled by the local counterterm
\begin{equation}
 \delta \mathcal{L}_{\rm CT}^{\beta}
=
\delta\mathcal{L}_{\beta\psi} + \delta \mathcal{L}_{\beta A}
+
\delta \mathcal{L}_{DF}
+
\delta\mathcal{L}_{3\gamma}\, .
\label{eq:CTbeta}
\end{equation}
Here ${\mathcal{L}}_{\beta\psi}$ denotes the complete fermionic sector at order $\beta$, while ${\mathcal{L}}_{\beta A}$ denotes the $\beta$-dependent deformation of the photon kinetic term. Both structures are already present in the original GUP Lagrangian in the massless limit \eqref{QEDG}. Explicitly,
\begin{align}
{\mathcal{L}}_{\beta\psi} = -i\beta\bar{\psi} \slashed{\partial}\square \psi + {\mathcal{L}}_{I}^\beta \equiv \beta \, \mathcal{O}_{\beta\psi} \, , \qquad {\mathcal{L}}_{\beta A} = \frac{\beta}{2} F^{\mu\nu} \Big(\square F_{\mu\nu} \Big) \equiv \beta \, \mathcal{O}_{\beta A} \, ,
\end{align}
where ${\mathcal{L}}_{I}^\beta$ contains the complete set of $\beta$-dependent fermion--photon interactions generated by the original GUP operator.
As mentioned previously, the ultraviolet divergences in the fermionic sector do not reproduce only the original GUP structure. In addition to ${\mathcal{O}}_{\beta\psi}$, two independent transverse operator structures, $\mathcal{O}_{DF}$ and $\mathcal{O}_{3\gamma}$, are required. Their corresponding counterterms are
\begin{gather} 
\delta{\mathcal{L}_{\beta\psi}} = \beta \, C_\epsilon \frac{5-4\xi}{2} \mathcal{O}_{\beta\psi} \, , \qquad \delta \mathcal{L}_{DF} = \beta \, C_\epsilon \frac{13+6\xi}{12} \mathcal{O}_{DF} \, , \notag \\  \delta \mathcal{L}_{3\gamma} = \beta \, C_\epsilon \left(\xi-\frac{1}{2}\right)\mathcal{O}_{3\gamma} \, .
\label{dLdeltas}
\end{gather}
with the common divergent coefficient
\begin{align}
C_\epsilon \equiv -
Q^2\frac{\alpha}{4\pi}
\frac{\mu^{2\epsilon}}{\hat{\epsilon}} \, .
\end{align}
An important consistency check is that the coefficient multiplying the original fermionic GUP structure is reproduced independently from Green functions with different numbers of external fields. In particular, the same coefficient $(5-4\xi)/2$ multiplying $\mathcal{O}_{\beta\psi}$ is recovered from the corresponding two-, three-, four-, and five-field sectors whenever that structure contributes. Likewise, the additional transverse structures associated with $\mathcal{O}_{DF}$ and $\mathcal{O}_{3\gamma}$ appear with the coefficients $(13+6\xi)/12$ and $\xi-1/2$, respectively. The fact that the same operator coefficients are consistently reconstructed from Green functions of different multiplicities provides a nontrivial check both of the loop calculation and of the gauge structure of the divergent effective action.

The situation in the purely photonic sector is different. The $\beta$-dependent contribution to the photon two-point function renormalizes the operator already present in the original Lagrangian,
\begin{align}
\delta\mathcal{L}_{\beta A} = (\delta_A + \delta_\beta^A) \frac{\beta}{2} F^{\mu\nu} \Big(\square F_{\mu\nu} \Big)
\end{align}
where $\delta_\beta^A$ denotes the renormalization counterterm associated with the $\beta$ coefficient in the photonic sector, while $\delta_A$ is the usual photon-field renormalization counterterm. We have written $\delta_\beta^A$ separately because, once operator mixing is allowed, there is no a priori reason for the renormalization of the photonic $\beta$ operator to coincide with that of the fermionic GUP sector.

Finally, the remaining counterterms are those of the standard massless QED Lagrangian,
\begin{align}
 \mathcal{L}_{\rm CT}^{\rm SM} &= i \delta_\psi \, \bar{\psi} \slashed{\partial}\psi  - \frac{\delta_A}{4} F^{\mu\nu}F_{\mu\nu} -(\delta_A/2 + \delta_\psi + \delta_e)eQA_\mu \bar\psi \gamma^\mu \psi - \frac{(\delta_A-\delta_\xi)}{2\xi} \left(\partial_\mu A^\mu \right)^2  .
\end{align}
As we shall observe in the following, at one loop and to first order in $\beta$, the ultraviolet-divergent effective action is not renormalized solely by a multiplicative redefinition of the original GUP parameter. Rather, the fermionic GUP operator generates a renormalization counterterm $\delta_\beta^\psi$, while the photonic higher-derivative operator carries its own renormalization counterterm $\delta_\beta^A$.

As usual, the bare
fields and parameters are related to the renormalized ones through
\begin{gather}
A_0^\mu = Z_A^{1/2} A^\mu \, ,
\qquad
\psi_0 = Z_\psi^{1/2}\psi \, ,
\qquad
\bar{\psi}_0 = Z_\psi^{1/2}\bar{\psi} \, ,
\nonumber\\
e_0 = Z_e e \, ,
\qquad
\xi_0 = Z_\xi \xi \, ,
\qquad
\beta_0^{A} = Z_\beta^{A}\beta_A \, ,
\qquad
\beta_0^{\psi} = Z_\beta^{\psi}\beta_\psi      \, .
\label{Zfunc}
\end{gather}
At one-loop order, the corresponding renormalization constants are
written as
\begin{align}
Z_A &= 1+\delta_A \, ,
&
Z_\psi &= 1+\delta_\psi \, ,
&
Z_e &= 1+\delta_e \, ,
\notag \\ 
Z_\xi &= 1+\delta_\xi \, ,
&
Z_\beta^{A} &= 1+\delta_\beta^{A} \, ,
&
Z_\beta^{\psi} &= 1+\delta_\beta^{\psi} \, .
\end{align}
Keeping only terms linear in the counterterms, the combinations entering
the renormalized Lagrangian become
\begin{align}
Z_e Z_A^{1/2} Z_\psi
=
1+\delta_e+\frac{\delta_A}{2}+\delta_\psi = 1+\delta_\psi \, ,  \qquad \frac{Z_A}{Z_\xi}
=
1+\delta_A-\delta_\xi = 1 \, ,
\end{align}
where ${\delta_A}/{2}+\delta_e = 0$ is the usual SM Ward identity and where, we have chosen $\delta_A=\delta_\xi$ as usual. The remaining combinations are given by
\begin{align}
Z_A^{1/2}
=
1+\frac{\delta_A}{2} 
\, ,
\qquad
Z_\beta^{A} Z_A
=
1+\delta_\beta^{A}+\delta_A \, .
\end{align}
These relations reproduce directly the combinations of counterterms
appearing in $\mathcal{L}_{\rm CT}^{\rm SM}$ and in
$\delta\mathcal{L}_{\beta A}$. The SM counterterms are, as usual, given by
\begin{align}
\delta_\psi = -\xi \, C_\epsilon \, , \qquad \delta_A = -\frac{4}{3} \, C_\epsilon \, , \qquad .
\end{align}

Let's now take a closer look at $\delta \mathcal{L}_{\beta\psi}$. 
Writing separately the different field-multiplicity sectors of the
fermionic GUP operator, we have
\begin{align}
\delta \mathcal{L}_{\beta\psi}
={}&
-i\beta \left(\delta_\beta^\psi+\delta_\psi\right)
\bar{\psi}\slashed{\partial}\square\psi
+
\left(
\delta_\beta^\psi+\delta_\psi+\delta_e+\frac{\delta_A}{2}
\right)
\mathcal{L}^\beta_{\psi\psi A}
\nonumber\\
&+
\left(
\delta_\beta^\psi+\delta_\psi+2\delta_e+\delta_A
\right)
\mathcal{L}^\beta_{\psi\psi AA} + 
\left(
\delta_\beta^\psi+\delta_\psi+3\delta_e
+\frac{3}{2}\delta_A
\right)
\mathcal{L}^\beta_{\psi\psi AAA} \, .
\label{eq:CTbetapsi-expanded}
\end{align}
Here, $\mathcal{O}^\beta_{\psi\psi A}$,
$\mathcal{O}^\beta_{\psi\psi AA}$ and
$\mathcal{O}^\beta_{\psi\psi AAA}$ denote, respectively, the
one-, two- and three-photon interaction structures generated by the
fermionic $\mathcal{O}(\beta)$ operator in the original interaction
Lagrangian $\mathcal{L}_{I}^{\beta}$. The corresponding powers of
$eQ$ are understood to be included in the definition of these
operators. The different combinations of renormalization constants in
Eq.~\eqref{eq:CTbetapsi-expanded} follow directly from \eqref{Zfunc}. Note that a term containing $n$ photon fields is multiplied by
\begin{align}
Z_\beta^\psi Z_\psi
\left(Z_e Z_A^{1/2}\right)^n \, .
\label{eq:ZnGUP}
\end{align}
Together with the Ward identity $Z_e Z_A^{1/2}=1$, at one-loop order, implies
\begin{align}
\delta_e+\frac{\delta_A}{2}= 2\delta_e+\delta_A= 3\delta_e
+\frac{3}{2}\delta_A = 0 \, .
\label{eq:Warddelta}
\end{align}
Therefore, the complete counterterm associated with the original
fermionic GUP operator can be written as
\begin{align}
\delta \mathcal{L}_{\beta\psi}
=
\left(\delta_\beta^\psi+\delta_\psi\right)
\mathcal{L}_{\beta\psi}
=
\beta \, \left(\delta_\beta^\psi+\delta_\psi\right)
\mathcal{O}_{\beta\psi} =  \beta \, C_\epsilon \frac{5-4\xi}{2} \mathcal{O}_{\beta\psi} 
 \, .
\label{eq:CTbetapsi-compact}
\end{align}
where we have used the deffinitions en \eqref{dLdeltas}. Therefore, the renormalization counterterm for $\beta$ from the fermionic sector is given by
\begin{align}
\delta_\beta^\psi = C_\epsilon \frac{5-4\xi}{2} - \delta_\psi =  C_\epsilon \frac{5-2\xi}{2} \, .
\end{align}

Taking into account all the previous considerations, at this point, the complete counter-term lagrangian, in terms of the corrspondign operators, reads
\begin{align}
\delta \mathcal{L}_{\rm CT}
&=
\beta \big(\delta_\beta^\psi+\delta_\psi \big)
\mathcal{O}_{\beta\psi} +  \beta \big(\delta_A + \delta_\beta^A \big)  \mathcal{O}_{\beta A}  + i \delta_\psi \, \bar{\psi} \slashed{\partial}\psi  - \frac{\delta_A}{4} F^{\mu\nu}F_{\mu\nu} - \delta_\psi eQA_\mu \bar\psi \gamma^\mu \psi  
 \notag \\
 & \qquad  + \beta \, C_\epsilon \frac{13+6\xi}{12} \mathcal{O}_{DF} + \beta \, C_\epsilon \left(\xi-\frac{1}{2}\right)\mathcal{O}_{3\gamma}
 \, , 
\label{LCT1}
\end{align}
where we must set
$\delta_\beta^A=-\delta_A$, since the $\mathcal O(\beta)$
contribution to the photon two-point function contains no additional
ultraviolet pole. Notice that this statement refers to the direct
renormalization of $\mathcal O_{\beta A}$ in the photon two-point
function and does not preclude contributions along the same physical
operator direction arising from the EOM reduction of counterterms
generated in other Green-function sectors. Accordingly, the set
\begin{align}
\left\{
\mathcal O_{\beta\psi},
\mathcal O_{\beta A},
\mathcal O_{DF},
\mathcal O_{3\gamma}
\right\}
\end{align}
together with the standard QED operators, provides a closed Green-function
basis for the ultraviolet divergences encountered in the two-point photon
and $2\psi+nA$ 1PI sectors considered above. However it turns out that the prevoious basis is redundant. Once the
off-shell renormalization has been carried out, integration by parts, the
Bianchi identity, and the leading-order equations of motion imply
nontrivial relations among the operators introduced above. In particular
(see Appendix A for an explicit derivation),
\begin{equation}
\mathcal{O}_{3\gamma}
\simeq
-\frac{1}{2}\mathcal{O}_{DF} \simeq  -\frac{1}{2} \mathcal{O}_{\beta A} ,
\end{equation}
where $\simeq$ denotes equality modulo total derivatives and
leading-order equations of motion. Furthermore,
\begin{equation}
\frac{13+6\xi}{12}\,
\mathcal{O}_{DF}
+
\left(\xi-\frac{1}{2}\right)
\mathcal{O}_{3\gamma}
\simeq
\frac{4}{3}\,
\mathcal{O}_{\beta A}.
\label{eq:physical_transverse_projection}
\end{equation}
which implies that the radiative corrections detected
in the $2\psi+nA$ sector do not generate a new physical operator
direction: after removal of redundant operators, they project back onto
the same physical direction already contained in the original GUP
deformation. 
After applying the EOM, the corresponding counterterm for the $\mathcal{O}_{\beta A}$ operator is given by
\begin{align}
\delta \mathcal{L}_{\beta A} =  \beta \, \frac{4}{3} C_\epsilon \, \mathcal{O}_{\beta A} = \beta \, \delta_A \, \mathcal{O}_{\beta A} \, . 
\end{align}
where we have identified $\delta_A$ with $4 C_\epsilon /3$. Therefore, after applying the EOM, and distinguishing between the two $\beta$ terms (the one corresponding to the ferionic sector and the one corresponding to the purely bosonic sector) the counter-term Lagrangian \eqref{LCT1} finally reduces to
\begin{align}
\delta \mathcal{L}_{\rm CT}
&=
\beta_\psi \big(\delta_\beta^\psi+\delta_\psi \big)
\mathcal{O}_{\beta\psi} +  \beta_A \, \delta_A \, \mathcal{O}_{\beta A}  + i \delta_\psi \, \bar{\psi} \slashed{\partial}\psi  - \frac{\delta_A}{4} F^{\mu\nu}F_{\mu\nu} - \delta_\psi eQA_\mu \bar\psi \gamma^\mu \psi 
 \, , 
\label{LCTfin}
\end{align}
and so, the physical basis for performing on-shell calculations at one loop level is given by
\begin{align}
\left\{
\mathcal O_{\beta\psi},
\mathcal O_{\beta A}
\right\}
\label{basis}
\end{align}
together with the standard QED operators.

\section{Four-fermion operators and complete basis}
\label{fourfermion}

When using the lowest-order Maxwell equation,
\begin{equation}
\partial_\nu F^{\nu\mu}
=
(eQ)\,\bar\psi\gamma^\mu\psi \, ,
\end{equation}
and introducing the four-fermion current-current operator
\begin{equation}
\mathcal{O}_{VV}
= \frac{1}{2}
\left(\bar\psi\gamma_\mu\psi\right)
\left(\bar\psi\gamma^\mu\psi\right),
\end{equation}
one obtains (see Appendix B for a complete demonstration)
\begin{equation}
\mathcal{O}_{\beta A}
\simeq
-2(eQ)^2\,\mathcal{O}_{VV},
\label{eq:GUP_VV_equivalence}
\end{equation}
However, $\mathcal{O}_{VV}$ should not be interpreted as an additional
physical direction with respect to $\mathcal{O}_{\beta A}$: the two
operators are simply different representatives of the same equivalence
class modulo the leading-order equations of motion. However, the previous
analysis does not exclude the possibility that genuinely new
four-fermion structures, inequivalent to $\mathcal{O}_{\beta A}$, are
generated radiatively. A complete closure test therefore requires the
explicit calculation of the four-fermion 1PI Green function. As shown
below, this additional sector is precisely where a new independent
axial-axial operator emerges.

\begin{figure}[t!]
\centering
\includegraphics[scale=0.4]{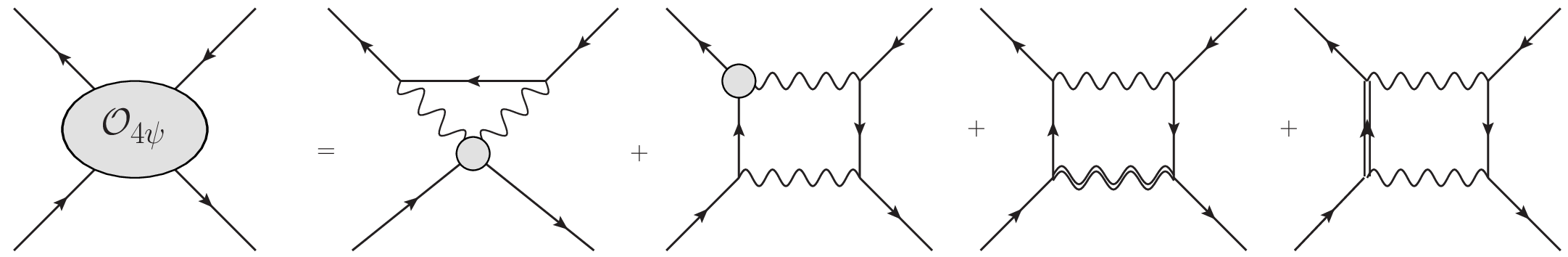}
\caption{Typical one-loop GUP contributions at $\mathcal{O}(\beta)$ to four-fermion vertex.} 
\label{ver4Psi_cor}
\end{figure} 

After performing the one-loop calculation of the four-fermion 1PI Green
function at first order in the GUP parameter $\beta$ (see Figure~\ref{ver4Psi_cor} for the topology classes involved in this process), an
additional ultraviolet structure is found. After extracting the ultraviolet pole in $D$ dimensions and only then
reducing the Dirac algebra to four dimensions, the divergent part can be
written as the sum of the two contributions
\begin{equation}
\Gamma^{(AA)}_{\epsilon}
=
-\frac{3i\beta (eQ)^4}{4\pi^2}
\frac{\mu^{2\epsilon}}{\hat\epsilon}
\left[
\left(\gamma^\mu\gamma^5\right)_{21}
\left(\gamma_\mu\gamma^5\right)_{43}
-
\left(\gamma^\mu\gamma^5\right)_{41}
\left(\gamma_\mu\gamma^5\right)_{23}
\right].
\label{eq:4psi_div} 
\end{equation}
and,
\begin{equation}
\Gamma_{\epsilon}^{(VV)}
=
-\frac{i\beta (eQ)^4}{6\pi^2}
\frac{\mu^{2\epsilon}}{\hat{\epsilon}}
\left[
\left(\gamma^\mu\right)_{21}
\left(\gamma_\mu\right)_{43}
-
\left(\gamma^\mu\right)_{41}
\left(\gamma_\mu\right)_{23}
\right].
\label{eq:4psi_VV_div}
\end{equation}
The relative minus sign between the two terms in \eqref{eq:4psi_div} is required by the
antisymmetry under the exchange of identical external fermions. The
three-gamma structures appearing in the direct calculation can be reduced
by means of the four-dimensional Chisholm identity. In particular,
\begin{equation}
\left(
\gamma^\mu\gamma^\nu\gamma^\rho
-
\gamma^\rho\gamma^\nu\gamma^\mu
\right)
\otimes
\gamma_\mu\gamma_\nu\gamma_\rho
=
12\,
\gamma^\sigma\gamma^5
\otimes
\gamma_\sigma\gamma^5 .
\end{equation}
It is therefore convenient to introduce the four-fermion operator
\begin{equation}
\mathcal{O}_{AA}
=
\frac{1}{2}
\left(
\bar{\psi}\gamma^\mu\gamma^5\psi
\right)
\left(
\bar{\psi}\gamma_\mu\gamma^5\psi
\right),
\label{eq:OAA}
\end{equation}
whose normalization has been chosen such that its four-fermion vertex is
\begin{equation}
i\left[
\left(\gamma^\mu\gamma^5\right)_{21}
\left(\gamma_\mu\gamma^5\right)_{43}
-
\left(\gamma^\mu\gamma^5\right)_{41}
\left(\gamma_\mu\gamma^5\right)_{23}
\right].
\end{equation}
The corresponding local counterterm is thus
\begin{equation}
\mathcal{L}^{(AA)}_{\rm CT}
=
+\frac{3\beta (eQ)^4}{4\pi^2}
\frac{\mu^{2\epsilon}}{\hat\epsilon}\,
\mathcal{O}_{AA},
\label{eq:CT_OAA}
\end{equation}
which is manifestly independent of the gauge parameter $\xi$.

Hence, two distinct gauge-independent four-fermion ultraviolet structures
are generated at one loop and at first order in $\beta$. The
vector--vector operator $\mathcal{O}_{VV}$ becomes manifest through the
projection of the off-shell counterterms already present in the
$2\psi+nA$ sector, whereas the axial--axial operator
$\mathcal{O}_{AA}$ is revealed only by the explicit calculation of the
four-fermion 1PI Green function.

This result has an important consequence for the operator analysis.
The Green functions containing two external fermions are sufficient to
reveal the off-shell structures
$\mathcal{O}_{\beta\psi}$,
$\mathcal{O}_{DF}$ and
$\mathcal{O}_{3\gamma}$, but they cannot probe genuinely new local
four-fermion interactions. The explicit four-fermion calculation
therefore does more than provide an independent consistency check: it
reveals the additional axial--axial structure
$\mathcal{O}_{AA}$ required by renormalization.

After projection of the off-shell Green-function basis by means of
integration by parts and the leading-order equations of motion, the
physical dimension-six operator space relevant at this order, the basis \eqref{basis} needs to be enlarged with an additional operator i.e.,  
\begin{equation}
\left\{
\mathcal{O}_{\beta\psi} ,
\mathcal{O}_{\beta A},
\mathcal{O}_{AA}
\right\},
\end{equation}

\section{One - loop renormalization and RGEs}
\label{rge}
We now discuss the renormalization-group evolution of the coefficients
multiplying the dimension-six operators. It is useful to distinguish
between the coefficients defined in the off-shell Green-function basis
and those associated with the reduced physical operator basis. Let us first consider the fermionic GUP coefficient. We recall the followign relations
\begin{equation}
\beta^{\psi}_0
=
Z_{\beta}^\psi \beta_\psi,
\qquad
Z_{\beta}^\psi
=
1+\delta_{\beta}^\psi = 1 + C_\epsilon\frac{5-2\xi}{2} \, .
\end{equation}
Accordingly, the corresponding one-loop evolution in the
Green-function basis is
\begin{equation}
\mu\frac{d\beta_\psi}{d\mu}
=
(5-2\xi)\,
Q^2\frac{\alpha}{4\pi}\,
\beta_\psi \, ,
\label{eq:RGEbetapsi}
\end{equation}
which gives, to one-loop accuracy,
\begin{equation}
\beta_\psi(\mu)
=
\beta(\Lambda)
\left[
1+
(5-2\xi)Q^2\frac{\alpha}{4\pi}
\ln\left(
\frac{\mu}{\Lambda}
\right)
\right]
+
\mathcal O(\alpha^2\beta),
\end{equation}
where we have chosen $\beta = \beta(\Lambda)$ as the tree-level parameter. We next consider the photonic GUP operator,
\begin{equation}
\mathcal O_{\beta A}
=
\frac{1}{2}
F_{\mu\nu}\Box F^{\mu\nu}.
\end{equation}
In the off-shell Green-function basis we have
\begin{equation}
\beta^{A}_0
=
Z_{\beta}^A \beta_A,
\qquad
Z_{\beta}^A
=
1+\delta_{\beta}^A.
\end{equation}
Since the $\mathcal O(\beta)$ contribution to the photon two-point
function contains no additional ultraviolet pole, we have found
\begin{equation}
\delta_{\beta}^A
+
\delta_A
=
0,
\end{equation}
thus, in the Green-function basis, the coefficient $\beta_A$ carries
a scale dependence compensating the photon-field renormalization.
The situation becomes simpler after eliminating the redundant
operators. Using
\begin{equation}
\mathcal O_{DF}
\simeq
\mathcal O_{\beta A},
\qquad
\mathcal O_{3\gamma}
\simeq
-\frac{1}{2}\mathcal O_{\beta A},
\end{equation}
the corresponding transverse counterterms satisfy
\begin{align}
&
C_\epsilon
\left[
\frac{13+6\xi}{12}\mathcal O_{DF}
+
\left(\xi-\frac12\right)\mathcal O_{3\gamma}
\right]
\simeq
\frac{4}{3}C_\epsilon
\mathcal O_{\beta A}
=
\delta_A\,
\mathcal O_{\beta A}.
\end{align}
Therefore, in the reduced physical basis the complete counterterm
associated with $\mathcal O_{\beta A}$ is
\begin{equation}
\delta\mathcal L_{\beta A}
=
\beta_A \, \delta_A\,
\mathcal O_{\beta A}.
\end{equation}
Since the factor $\delta_A$ is already generated by the
renormalization of the two photon fields contained in
$\mathcal O_{\beta A}$, no additional multiplicative renormalization
of the physical Wilson coefficient is required:
\begin{equation}
Z_{\beta}^A = 1 \, ,
\end{equation}
at one loop and to first order in $\beta$. Thus, although the coefficient of $\mathcal O_{\beta A}$ appears to
run in the redundant Green-function basis, its physical coefficient
does not possess an anomalous dimension at this order.

The four-fermion sector introduces a qualitatively different effect.
The one-loop calculation generates the additional physical operator
\begin{equation}
\mathcal O_{AA}
=
\frac12
\left(
\bar\psi\gamma^\mu\gamma^5\psi
\right)
\left(
\bar\psi\gamma_\mu\gamma^5\psi
\right),
\end{equation}
whose coefficient will be denoted by $C_{AA}$. The physical
dimension-six Lagrangian can therefore be written as
\begin{equation}
\mathcal L_{6}
= \beta_\psi\,\mathcal O_{\beta \psi} + 
\beta_A\,\mathcal O_{\beta A}
+
C_{AA}\,\mathcal O_{AA}.
\end{equation}
At the matching scale $\Lambda$ we impose
\begin{equation}
\beta_A =  \beta_\psi(\Lambda) \equiv \beta,
\qquad
C_{AA}(\Lambda)=0.
\end{equation}
Using the one-loop mixing coefficient obtained from the four-fermion
divergence, the renormalization-group equations take the form
\begin{equation}
\mu\frac{dC_{AA}}{d\mu}
=
-\frac{3(eQ)^4}{2\pi^2}\,
\beta
.
\label{eq:RGECAA}
\end{equation}
whose solution with the above matching conditions is and
\begin{align}
C_{AA}(\mu)
&=
-\frac{3\beta (eQ)^4}{2\pi^2}
\ln\left(
\frac{\mu}{\Lambda}
\right) \, .
\end{align}

\section{Discussion and conclusions}

The results presented above have a more general consequence for the
quantum consistency of the GUP deformation. The Green functions
containing two external fermions are sufficient to reveal the off-shell
structures $\mathcal{O}_{\beta\psi}$, $\mathcal{O}_{DF}$ and
$\mathcal{O}_{3\gamma}$, but they cannot probe genuinely new local
four-fermion interactions. The explicit four-fermion calculation
therefore does more than provide an independent consistency check: it
reveals the axial--axial operator $\mathcal{O}_{AA}$ as a genuinely
radiatively generated physical structure which cannot be inferred from
the classical equations of motion alone.

After projection of the off-shell counterterms by means of integration
by parts and the leading-order equations of motion, the combination of
$\mathcal{O}_{DF}$ and $\mathcal{O}_{3\gamma}$ does not generate a
second independent physical direction. Rather, it projects back onto
the original photonic GUP operator $\mathcal{O}_{\beta A}$. Equivalently,
introducing
\begin{equation}
\mathcal{O}_{VV}
=
\frac{1}{2}
\left(\bar\psi\gamma_\mu\psi\right)
\left(\bar\psi\gamma^\mu\psi\right),
\end{equation}
the leading-order Maxwell equation gives
\begin{equation}
\mathcal{O}_{\beta A}
\simeq
-2(eQ)^2\,\mathcal{O}_{VV}.
\end{equation}
Thus, $\mathcal{O}_{VV}$ and $\mathcal{O}_{\beta A}$ are not two
independent physical operators, but two different representatives of
the same equivalence class modulo the leading-order equations of
motion.

Remarkably, the coefficient multiplying this original physical GUP
direction is independent of the covariant gauge parameter after the
projection, while the coefficient of the independently generated
$\mathcal{O}_{AA}$ operator is also gauge independent. The cancellation
of the nontrivial $\xi$ dependence present in the intermediate off-shell
Green functions therefore provides a nontrivial consistency check of
the separation between redundant Green-function operators and physical
ultraviolet structures.

The physical ultraviolet divergence may be written in a basis adapted
to the original GUP deformation as
\begin{equation}
\mathcal{L}_{\rm div}^{\rm phys}
\simeq
\frac{\beta(eQ)^2}{12\pi^2}
\frac{\mu^{2\epsilon}}{\hat{\epsilon}}\,
\mathcal{O}_{\beta A}
-
\frac{3\beta(eQ)^4}{4\pi^2}
\frac{\mu^{2\epsilon}}{\hat{\epsilon}}\,
\mathcal{O}_{AA}.
\label{eq:physical_div_GUP_basis}
\end{equation}
Equivalently, using the Maxwell equation, the same physical divergence
can be represented entirely in the four-fermion basis as
\begin{equation}
\mathcal{L}_{\rm div}^{\rm phys}
\simeq
-\frac{\beta(eQ)^4}{\pi^2}
\frac{\mu^{2\epsilon}}{\hat{\epsilon}}
\left[
\frac{1}{6}\mathcal{O}_{VV}
+
\frac{3}{4}\mathcal{O}_{AA}
\right].
\label{eq:physical_div_4f_basis}
\end{equation}
Hence the physical dimension-six operator space required at this order
is three-dimensional and may be represented equivalently as
\begin{equation}
\left\{
\mathcal{O}_{\beta \psi},
\mathcal{O}_{\beta A},
\mathcal{O}_{AA}
\right\}
\qquad\Longleftrightarrow\qquad
\left\{
\mathcal{O}_{\beta \psi},
\mathcal{O}_{VV},
\mathcal{O}_{AA}
\right\}.
\end{equation}
The important point is that only one of these directions is genuinely
new. The original physical GUP direction is preserved by the
one-loop corrections, whereas quantum fluctuations generate the
additional independent axial--axial interaction
$\mathcal{O}_{AA}$. The original GUP deformation is therefore not
multiplicatively closed under renormalization, but its quantum
completion at this order requires one additional physical operator.

The four-fermion representation makes the chiral content of this new
quantum structure particularly transparent. Defining
\begin{equation}
\mathcal{O}_{+}
=
\mathcal{O}_{VV}+\mathcal{O}_{AA},
\qquad
\mathcal{O}_{-}
=
\mathcal{O}_{VV}-\mathcal{O}_{AA},
\end{equation}
the ultraviolet divergence becomes
\begin{equation}
\mathcal{L}_{\rm div}^{\rm phys}
=
-\frac{\beta(eQ)^4}{24\pi^2}
\frac{\mu^{2\epsilon}}{\hat{\epsilon}}
\left(
11\,\mathcal{O}_{+}
-
7\,\mathcal{O}_{-}
\right).
\end{equation}
In terms of the right- and left-handed currents,
\begin{equation}
J^\mu_{R,L}
=
\bar\psi_{R,L}\gamma^\mu\psi_{R,L},
\end{equation}
one equivalently finds
\begin{equation}
\frac{1}{6}\mathcal{O}_{VV}
+
\frac{3}{4}\mathcal{O}_{AA}
=
\frac{11}{24}
\left(
J_R^\mu J_{R\mu}
+
J_L^\mu J_{L\mu}
\right)
-
\frac{7}{12}
J_R^\mu J_{L\mu}.
\end{equation}
Thus, the radiatively generated component distinguishes same- and
opposite-chirality fermion configurations while preserving parity.
This should not be interpreted as a rotation of the original GUP
direction into two new operator directions. Rather, the original
vector-current equivalence class remains present, while the quantum
theory develops an additional axial--axial component.

A particularly interesting feature of the radiatively generated
$\mathcal{O}_{AA}$ operator is that it does not signal parity or
$CP$ violation. Although it is constructed from axial currents, their
Lorentz contraction is parity even. Instead, it introduces a genuine
spin- and chirality-dependent contact interaction. In the
non-relativistic limit, the axial--axial contribution corresponds to a
short-range spin-dependent interaction and therefore represents a
qualitatively new physical effect which is absent from the original
classical GUP deformation.

A noteworthy aspect of this result is the form of the genuinely new
operator generated by the quantum corrections. Four-fermion interactions
constructed from axial currents arise naturally in gravitational theories
with spacetime torsion. In the Einstein--Cartan formulation, the torsional
part of the spin connection is non-propagating at the lowest derivative
order and can be eliminated algebraically. In the presence of minimally
coupled Dirac fermions, this procedure generates an effective local
axial--axial interaction of the form
\begin{equation}
\Delta\mathcal{L}_{\rm torsion}
\propto
\left(
\bar{\psi}\gamma^\mu\gamma^5\psi
\right)
\left(
\bar{\psi}\gamma_\mu\gamma^5\psi
\right),
\end{equation}
which has precisely the same Lorentz and Dirac structure as the
$\mathcal{O}_{AA}$ operator generated here
\cite{FreidelMinicTakeuchi2005,DiakonovTumanovVladimirov2011}.

This correspondence is particularly suggestive because
$\mathcal{O}_{AA}$ is absent from the physical tree-level GUP
deformation and appears only after quantum corrections are included.
In the present theory, however, its origin is entirely different:
no torsional degree of freedom has been introduced, and the operator
is generated radiatively by the GUP-modified electromagnetic dynamics.
The correspondence should therefore be understood as an equivalence
at the level of the low-energy operator structure, rather than as
evidence that spacetime torsion is dynamically generated in GUP-QED.

There is also an important difference between the two mechanisms.
For minimally coupled fermions in Einstein--Cartan gravity, elimination
of torsion produces the axial--axial direction, whereas in the present
case the quantum theory retains the original physical GUP direction,
which may equivalently be represented by the vector--vector operator,
\begin{equation}
\mathcal{O}_{\beta A}
\simeq
-2(eQ)^2\mathcal{O}_{VV},
\end{equation}
and generates $\mathcal{O}_{AA}$ in addition. Thus the physical
one-loop structure of GUP-QED can be represented schematically as
\begin{equation}
\mathcal{O}_{\beta A}
\quad\longrightarrow\quad
\left\{
\mathcal{O}_{\beta A},
\mathcal{O}_{AA}
\right\},
\end{equation}
or, equivalently, in the contact-interaction basis as
\begin{equation}
\mathcal{O}_{VV}
\quad\longrightarrow\quad
\left\{
\mathcal{O}_{VV},
\mathcal{O}_{AA}
\right\}.
\end{equation}
The axial--axial interaction is therefore the genuinely new physical
direction generated by the quantum completion of the GUP deformation.

The comparison becomes even more informative in more general theories
with torsion. Non-minimal fermion--torsion couplings may generate
axial--vector interactions and consequently parity-violating effects,
while general low-energy torsion effective actions may contain
vector--vector, axial--vector and axial--axial four-fermion structures
\cite{FreidelMinicTakeuchi2005,DiakonovTumanovVladimirov2011}.
By contrast, the one-loop GUP deformation considered here generates no
axial--vector interaction: the new $\mathcal{O}_{AA}$ contribution is
parity even. The resulting pattern of effective operators therefore
provides, at least in principle, a way of distinguishing the GUP-induced
interaction from more general torsion scenarios.

It is therefore remarkable that a deformation originally motivated by
short-distance modifications of quantum mechanics develops, through
ordinary quantum loops, a spin-dependent contact interaction belonging
to the same operator class that characterizes fermionic couplings to
spacetime torsion. This does not establish a dynamical connection between
the two frameworks, but it suggests that apparently different
short-distance completions may become partially degenerate when described
at the level of their low-energy effective operators.

\begin{appendix}

\section{EOM equivalence of the $\mathcal{O}_{\beta A}$, $\mathcal{O}_{DF}$ and $\mathcal{O}_{3\gamma}$ operators}

\label{EOMrelations}

In this appendix we explicitly derive the operator relations used in the
main text. We define
\begin{align}
\mathcal O_{DF}
&= -(eQ)\,
   (\bar\psi\gamma_\mu\psi)\,
   \partial_\nu F^{\nu\mu},
\\
\mathcal O_{3\gamma}
&= -\frac{eQ}{4}\,
   \bar\psi\gamma^{\mu\nu\rho}
   F_{\mu\nu}
   \overleftrightarrow D_\rho\psi,
\\
\mathcal O_{\beta A}
&= \frac12 F_{\mu\nu}\Box F^{\mu\nu}.
\end{align}
Throughout this appendix, $\simeq$ denotes equality modulo total
derivatives and the leading-order equations of motion. Since the
dimension-six Lagrangian is retained only to first order in $\beta$,
the use of the lowest-order QED equations of motion is sufficient,
as higher-order corrections would contribute only at
$\mathcal O(\beta^2)$.
For massless QED, the relevant equations of motion are
\begin{equation}
\slashed D\psi=0,
\qquad
\bar\psi\overleftarrow{\slashed D}=0,
\qquad
\partial_\nu F^{\nu\mu}
=(eQ)\,\bar\psi\gamma^\mu\psi.
\label{eq:LOEOM_app}
\end{equation}
We first consider $\mathcal O_{3\gamma}$. Using
\begin{equation}
\gamma^{\mu\nu\rho}
=
\gamma^\mu\gamma^\nu\gamma^\rho
-\eta^{\mu\nu}\gamma^\rho
+\eta^{\mu\rho}\gamma^\nu
-\eta^{\nu\rho}\gamma^\mu ,
\end{equation}
together with the fermion equations of motion and the antisymmetry of
$F_{\mu\nu}$, one obtains
\begin{align}
\bar\psi\gamma^{\mu\nu\rho}F_{\mu\nu}D_\rho\psi
&\simeq
2F^{\rho}{}_{\mu}\,
\bar\psi\gamma^\mu D_\rho\psi,
\\
(D_\rho\bar\psi)\gamma^{\mu\nu\rho}F_{\mu\nu}\psi
&\simeq
-2F^{\rho}{}_{\mu}\,
(D_\rho\bar\psi)\gamma^\mu\psi.
\end{align}
Therefore,
\begin{align}
\bar\psi\gamma^{\mu\nu\rho}F_{\mu\nu}
\overleftrightarrow D_\rho\psi
&\simeq
2F^{\rho}{}_{\mu}
\left[
\bar\psi\gamma^\mu D_\rho\psi
+
(D_\rho\bar\psi)\gamma^\mu\psi
\right]
\nonumber\\
&=
2F^{\rho}{}_{\mu}
\partial_\rho
(\bar\psi\gamma^\mu\psi)
\nonumber\\
&\simeq
-2(\partial_\rho F^{\rho}{}_{\mu})
(\bar\psi\gamma^\mu\psi).
\end{align}
It follows immediately that
\begin{equation}
\mathcal O_{3\gamma}
\simeq
\frac{eQ}{2}
(\bar\psi\gamma_\mu\psi)
\partial_\nu F^{\nu\mu}
=
-\frac12\,\mathcal O_{DF}.
\label{eq:O3_ODF_app}
\end{equation}
We next relate $\mathcal O_{DF}$ to the photonic GUP operator.
Integrating by parts gives
\begin{equation}
\mathcal O_{\beta A}
=
\frac12 F_{\mu\nu}\Box F^{\mu\nu}
\simeq
-\frac12
(\partial_\rho F_{\mu\nu})
(\partial^\rho F^{\mu\nu}).
\label{eq:ObetaA_IBP}
\end{equation}
The Bianchi identity,
\begin{equation}
\partial_\rho F_{\mu\nu}
+\partial_\mu F_{\nu\rho}
+\partial_\nu F_{\rho\mu}=0,
\end{equation}
implies
\begin{align}
(\partial_\rho F_{\mu\nu})
(\partial^\rho F^{\mu\nu})
&=
-2(\partial_\mu F_{\nu\rho})
   (\partial^\rho F^{\mu\nu})
\nonumber\\
&\simeq
2(\partial_\mu F^{\mu\nu})
 (\partial^\rho F_{\rho\nu}).
\end{align}
Hence,
\begin{equation}
\mathcal O_{\beta A}
\simeq
-(\partial_\mu F^{\mu\nu})
 (\partial^\rho F_{\rho\nu}).
\label{eq:Obeta_divF}
\end{equation}
Using the Maxwell equation in Eq.~\eqref{eq:LOEOM_app},
\begin{align}
\mathcal O_{\beta A}
&\simeq
-(eQ)(\bar\psi\gamma^\nu\psi)
\partial^\rho F_{\rho\nu}
\nonumber\\
&=
\mathcal O_{DF}.
\label{eq:ODF_ObetaA_app}
\end{align}
Combining Eqs.~\eqref{eq:O3_ODF_app} and
\eqref{eq:ODF_ObetaA_app}, one finally obtains
\begin{equation}
\mathcal O_{3\gamma}
\simeq
-\frac12\,\mathcal O_{DF}
\simeq
-\frac12\,\mathcal O_{\beta A}.
\label{eq:operator_equivalence_app}
\end{equation}

These relations hold in the reduced EFT operator basis, modulo total
derivatives and the leading-order equations of motion. The three
operators therefore correspond to the same physical operator direction,
although they remain distinct structures in the off-shell
Green-function basis used for renormalization.

\section{EOM equivalence of the $\mathcal{O}_{\beta A}$ and $\mathcal{O}_{VV}$ operators}
\label{EOM2}

We finally show the equivalence between the photonic GUP operator and
the vector four-fermion operator. Recall that
\begin{equation}
\mathcal O_{VV}
=
\frac{1}{2}
\left(\bar\psi\gamma_\mu\psi\right)
\left(\bar\psi\gamma^\mu\psi\right).
\end{equation}
We have previuosly shown that 
\begin{equation}
\mathcal O_{\beta A}
\simeq
-
(\partial_\mu F^{\mu\nu})
(\partial^\rho F_{\rho\nu}).
\label{eq:ObetaA_divF_app}
\end{equation}
The leading-order Maxwell equation,
\begin{equation}
\partial_\mu F^{\mu\nu}
=
(eQ)\,\bar\psi\gamma^\nu\psi,
\end{equation}
then yields
\begin{align}
\mathcal O_{\beta A}
&\simeq
-(eQ)^2
\left(\bar\psi\gamma_\nu\psi\right)
\left(\bar\psi\gamma^\nu\psi\right)
\nonumber\\
&=
-2(eQ)^2\,\mathcal O_{VV}.
\end{align}
Hence,
\begin{equation}
\mathcal O_{\beta A}
\simeq
-2(eQ)^2\,\mathcal O_{VV}.
\label{eq:ObetaA_OVV_app}
\end{equation}

\end{appendix}


\bibliographystyle{JHEP}
\bibliography{cite}

\end{document}